\documentclass[twocolumn,showpacs,preprintnumbers,amsmath,amssymb]{revtex4}
\usepackage{graphicx}
\usepackage{dcolumn}
\usepackage{bm}
\usepackage{multirow}

\begin{document}

\newcommand{\PL}[3]{Phys. Lett. {\bf #1}, #2 (#3)}
\newcommand{\PRL}[3]{Phys. Rev. Lett. {\bf #1}, #2 (#3)}
\newcommand{\NP}[3]{Nucl. Phys. {\bf #1}, #2 (#3)}
\newcommand{\PR}[3]{Phys. Rev. {\bf #1}, #2 (#3)}
\newcommand{\PTP}[3]{Prog. Theor. Phys. {\bf #1}, #2 (#3)}
\newcommand{\andvol}[3]{{\bf #1}, #2 (#3)}
\newcommand{\kbar}{$\bar{\rm K}$}
\newcommand{\km}{K$^-$}


\title{Invariant-Mass Spectroscopy for Condensed Single- and Double-\kbar~Nuclear Clusters\\ 
 to be Formed as Residues in Relativistic Heavy-Ion Collisions}

\author{Toshimitsu Yamazaki$^a$, Akinobu Dot$\acute{\rm e}$$^b$ and Yoshinori Akaishi$^b$, }
\address{$^a$RI Beam Science Laboratory, RIKEN, Wako,
Saitama 351-0198, Japan,}
\address{$^b$Institute of Particle and Nuclear Studies, KEK, Tsukuba, Ibaraki 305-0801, Japan}

\date{\today}

\begin{abstract}
Using a phenomenological
\={K}N interaction combined with the method of Antisymmetrized Molecular Dynamics., we predict that few-body double-\kbar~nuclei,  such as  ppK$^-$K$^-$ and ppnK$^-$K$^-$, as well as single-\kbar~nuclei, are tightly bound compact systems with large binding energies and ultra-high nucleon densities.  We point out that these \kbar~nuclear clusters can be produced as residual fragments in relativistic heavy-ion collisions  and that their invariant masses can be reconstructed from their decay particles.  
\end{abstract}




\maketitle

\noindent
{\bf 1. Introduction}\\

Recently, exotic nuclear systems involving a $\bar{\rm K}$ (K$^-$ or 
\={K}$^0$) as a constituent
have been investigated theoretically \cite{Akaishi:02,Yamazaki:02,Dote:02,Dote:03} based on phenomenologically constructed \={K}N
interactions (hereafter referred to as AY), which reproduce low-energy \={K}N
scattering data \cite{Martin},   kaonic hydrogen atom data \cite{Iwasaki}
and  the binding energy and decay width of $\Lambda$(1405).    These interactions are consistent with a prediction based on a chiral SU(3) effective Lagrangian \cite{Weise:96} and with a recent experimental indication on decreased in-medium K$^-$ mass from subthreshold nuclear reactions \cite{Schroter:94,Laue:99}. They are characterized
by a strongly attractive $I=0$ part,  causing drastic shrinkage of \kbar-bound nuclei and increasing the binding energies in proton-rich nuclei. 
The predicted bound states in  ppnK$^-$,
ppnnK$^-$ and $^8$BeK$^-$ lie below the $\Sigma\pi$ emission threshold,  and thus are expected to have narrow decay widths. These few-body treatments have been further extended to more complex systems by the method of Antisymmetrized
Molecular Dynamics (AMD) \cite{Dote:02,Dote:03}, which is now capable of calculating the structure with density distributions of individual constituent particles in  an {\it ab initio} way without any {\it a priori} assumption on the structure. 

The predicted \kbar~bound states have enormous nucleon densities ($\rho$) at the center, 4-9 times as much as the normal nuclear density ($\rho_0 = 0.17$ fm$^{-3}$), with large binding energies ($E_K \approx 100$ MeV). Such compact systems can be called ``\kbar~ nuclear clusters". Since the predicted nucleon densities very much exceed the nucleon compaction limit, $\rho_c \approx 1/v_N \approx 2.3 \rho_0$, with $v_N \approx 2.5$ fm$^3$ being the nucleon volume, these \kbar~clusters may be in deconfined quark-gluon states, which can better be named ``$s$-quark nuclear clusters". Interesting  questions naturally arise: how about the structure of double-\kbar~nuclei and how can they be produced and identified. In the present paper we report on the results of our calculations on the structure of the simplest systems, ppK$^-$K$^-$ and  ppnK$^-$K$^-$, and then propose to identify \kbar~clusters as residues (``\kbar~fragments") after relativistic heavy-ion reactions. This method, {\it decay-channel spectroscopy}, is to reconstruct invariant-mass spectra of decay particles of \kbar~clusters, in contrast to {\it formation-channel spectroscopy} to use direct reactions, such as (K$^-$, n) \cite{Akaishi:02,Kishimoto:99} and (K$^-, \pi^-$) \cite{Yamazaki:02}. \\  

\noindent
{\bf 2. Double \kbar~clusters}\\

\noindent
{\bf 2.1 ppK$^-$K$^-$}\\

\begin{figure}[htb]
\includegraphics[width=0.5\textwidth]{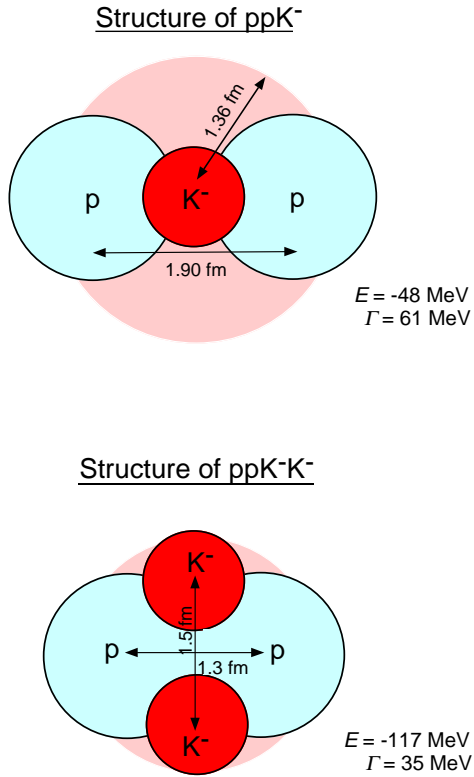}%
\caption{\label{fig:ppKK-dango}Schematic structure diagrams for the calculated ppK$^-$ and ppK$^-$K$^-$ nuclei. The rms radius of K$^-$ and rms inter-nucleon and inter-\kbar~distances are shown. }
\end{figure}

We applied the same theoretical treatments as given in \cite{Akaishi:02, Dote:02} to double-\kbar~systems. We used the 
Tamagaki potential (OPEG) \cite{Tamagaki} as a bare NN
interaction and the
AY \={K}N interaction as a bare \={K}N interaction, whereas we neglected the \kbar-\kbar~interaction simply because of a lack of information. 
We show the result of a variational calculation in  Fig.~\ref{fig:ppKK-dango}. The hitherto untouched ppK$^-$ system was predicted in a previous paper to be bound with a binding energy  ($E_K$ = 48 MeV) and a width ($\Gamma_K$ = 61 MeV) \cite{Yamazaki:02}. The p-p rms distance is 1.90 fm, close to the normal inter-nucleon distance. In the ppK$^-$K$^-$ system, on the other hand, the binding energy and width were  calculated to be $E_K$ = 117 MeV and $\Gamma_K$ = 35 MeV, and  the p-p rms distance  is very much reduced to 1.3 fm. Thus, the addition of a 
\kbar~increases the binding energy and the nucleon density. Since these bound states lie above the $\Sigma \pi$ emission threshold, their widths are dominated by the main decay channel (K$^-$p $\rightarrow \Sigma \pi$). \\

\noindent
{\bf 2.2 ppnK$^-$K$^-$ and pppnK$^-$K$^-$}\\

\begin{figure}[htb]
\includegraphics[width=0.25\textwidth]{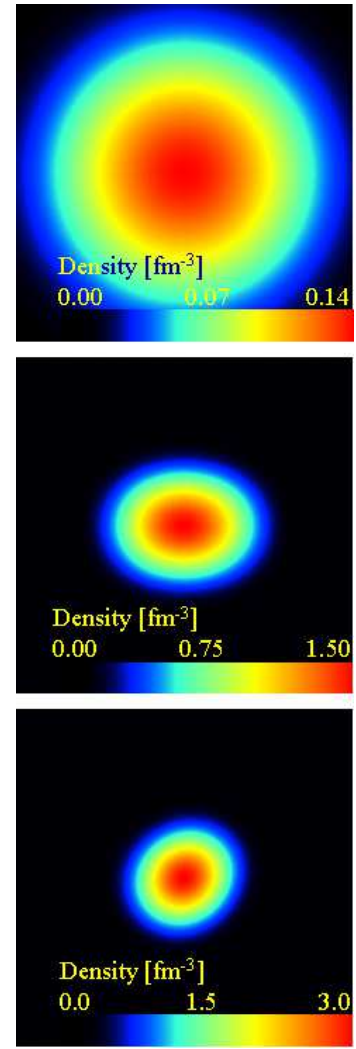}
\vspace{3mm}
\caption{\label{fig:ppnKK}Calculated density contours of ppn, ppnK$^-$ and ppnK$^-$K$^-$.}
\end{figure}

It was already predicted that the ppnK$^-$ system has a much stronger binding and a much higher density than ppK$^-$, indicating that the addition of a neutron further strengthens the binding of the system. Thus, it is interesting to investigate the ppnK$^-$K$^-$ system. We 
  constructed the effective NN-central force and the \={K}N force
with the $G$-matrix method, and carried out an AMD calculation of ppnK$^-$K$^-$. We found that the double-\kbar~cluster (ppnK$^-$K$^-$) is indeed more tightly bound than the single-\kbar~cluster (ppnK$^-$), as shown in Fig.~\ref{fig:ppnKK}, where we present the density contours of $^3$He, ppnK$^-$ and ppnK$^-$K$^-$. The central nucleon reaches $\rho (0) \sim 3 $ fm$^{-3}$. The pppnK$^-$K$^-$ system is shown to be bound even deeper. We summarize these results in Table~\ref{tab:summary} together with the results on single-\kbar~clusters \cite{Dote:02,Dote:03}. $\Gamma_{\rm K}$, the width for decaying to $\Lambda\pi$ and $\Sigma\pi$,
   was evaluated by calculating the expectation value
of the imaginary potential
contained in the effective  AY \={K}N interaction with the wave function
obtained by the AMD calculation. No additional widths of other origins are taken into account at this stage.\\

\noindent
{\bf 2.3 Possible suppression of the direct formation and decay of \kbar~clusters}\\

The compact \kbar~clusters predicted here are very different from ordinary nuclei in many respects. Their densities at the center are extremely high in view of the expected boundary between the hadron phase and the quark-gluon phase \cite{Hatsuda,Weise:93,CSC,phase-diagram2}. Thus, their structure can most likely  be described in terms of deconfined quarks, rather than of ``{\it nucleons + K}$^-$".  Although the nuclear structure of such {\it cold and dense} strange nuclei, possibly in a quark-gluon phase, has not been touched theoretically,  we expect that their decays to hadrons may be suppressed because of the need to rearrange of quarks and gluons into hadrons. On one hand, this would be a welcome feature, because the possible suppression of decays of \kbar~clusters favours the discreteness of these bound states for better spectroscopic observation. On the other hand, we anticipate that the same mechanism would also reduce the formation probability of these clusters via direct reactions on normal nuclear targets, such as (K$^-, \pi^-$) and ($\pi^+$, K$^+$) reactions for single-\kbar~nuclei \cite{Yamazaki:02}.
 Few-body double-\kbar~clusters  can in principle be produced by $\Delta S=-2$ direct reactions: d(K$^-$, K$_S^0$)ppK$^-$K$^-$ and $^3$He(K$^-$, K$^+$)ppnK$^-$K$^-$. When the formation of such a bound state is suppressed,
 no visible peak may be shown above a quasi-free background of p(K$^-$,K$^+)\Xi^-$ in an inclusive spectrum. This would be a serious problem.

\begin{table}
\caption{\label{tab:summary}
Summary of predicted \kbar~clusters. $M$: total mass [MeV]. $E_K$: total binding energy [MeV]. $\Gamma_K$: decay width [MeV]. 
$\rho (0)$: nucleon density at the center of the system [fm$^{-3}$]. $R_{\rm rms}$:
  root-mean-square radius of the nucleon system [fm]. $k_p$ and $k_K$: rms internal momenta [fm$^{-1}$] of p and K$^-$, respectively .
}
\begin{ruledtabular}
\begin{tabular}{l|ccccccc}
\kbar~cluster    & $Mc^2$ & $E_K$  &  $\Gamma_K$ & $\rho (0)$ & $R_{\rm rms}$ & $k_p$  & $k_K$ \\
     & [MeV] & [MeV] & [MeV]    & [fm$^{-3}$] & [fm] & [fm$^{-1}$] &  [fm$^{-1}$]  \\
\hline
$pK^-$ &1407 & 27   &   40  & 0.59     &   0.45       &  1.37   & 1.37  \\
$ppK^-$ &2322   & 48   &   61   &  0.52     &   0.99       & 1.49   & 1.18   \\  $pppK^-$ &3211 & 97   &   13 & 1.56    &    0.81  &      &     \\   
$ppnK^-$ &3192  & 118 &   21 & 1.50   &   0.72  &     &    \\  
$ppppK^-$ &4171& 75   &  162  & 1.68        & 0.95         &     &     \\ 
$pppnK^-$ &4135 &113  &   26   & 1.29 & 0.97 &     &     \\  
$ppnnK^-$ &4135 & 114  &  34   &         & 1.12  &     &     \\  
\hline
$ppK^-K^-$ &2747 & 117  & 35 &  &  &    &      \\
$ppnK^-K^-$&3582  & 221  & 37 & 2.97  & 0.69  &    &     \\
$pppnK^-K^-$ &4511  & 230  & 61 & 2.33  & 0.73  &    &      \\
\end{tabular}
\end{ruledtabular}
\end{table}

\vspace{0.5cm}
\noindent
{\bf 3. \kbar~clusters as residues in heavy-ion collisions}\\

Now, we point out that \kbar~clusters may be found as residues of relativistic  heavy-ion reactions, where K$^-$ mesons and $\Lambda$ hyperons are produced abundantly \cite{Laue:99,FOPI}. Usually, these strange particles are used as probes to study the size and temperature of fireballs produced in heavy-ion collisions. Here, we present a totally different view, namely, we propose to search for single-\kbar~and  double-\kbar~clusters as residues of hot and dense fireballs, since the probability of forming strongly bound \kbar~clusters is expected to be rather high. The dense medium provided in heavy-ion collisions should enhance such \kbar~cluster productions. Furthermore,  once  a  \kbar~cluster having a binding energy of $\sim 100$ MeV is produced in a chaotic nuclear medium, its tight binding  will make its dissociation difficult even at a high temperature of $50 \sim 100$ MeV. Thus, 
\kbar~clusters, once created, tend to survive through collisions, and escape in the freeze-out phase. They ultimately decay via their own  decay modes, from which the invariant masses of the parent \kbar~clusters may be reconstructed. \\

\begin{figure}[htb]
\vspace{1cm}
\includegraphics[width=0.4\textwidth]{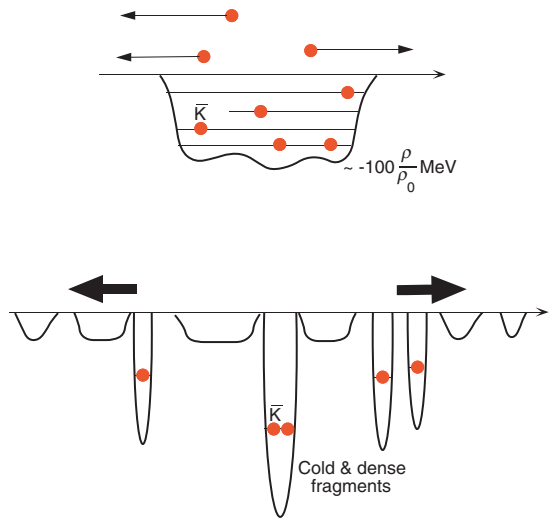}
\vspace{0cm}
   \caption{\label{fig:HI-continuum} (Upper) Chaotic continuum in heavy-ion reactions, where in-medium K$^-$ mesons are produced randomly. Some of K$^{-}$'s escape from the nuclear region as free K$^-$ mesons, whereas the others may form tightly bound \kbar~clusters. (Lower) Deep self-trapping potentials for \kbar~clusters are produced intermittently,  where K$^{-} $ and a few nucleons encounter.}
\end{figure}

In central collisions of relativistic heavy ions, a dense and hot fireball is produced, and as the fireball expands, they reach a ``freeze-out" phase, in which the produced hadrons are expected to be in thermal equilibrium. 
Recently, it is shown that particle emission data are well accounted for by a thermal equilibrium model in terms of a temperature ($T_f$) and a baryon chemical potential ($\mu_B$) as parameters 
\cite{Munzinger:95,Cleymans:98,Stock:99}. In the following we consider various steps toward the formation and decay of \kbar~clusters.\\

\noindent
{\it i) Abundant production of K$^{-}$'s in heavy-ion reactions}

 K$^-$ mesons are abundantly produced even in subthreshold nuclear reactions \cite{Schroter:94,Laue:99}. This phenomenon is interpreted as being due to the decreased K$^-$ mass in the nuclear medium, which is caused by a strong attraction between K$^-$ and p. They are embedded in an attractive nuclear potential (the mass of K$^{-} $ is effectively reduced), as shown in Fig.~\ref{fig:HI-continuum} (upper)  and continue to undergo collisions with nucleons.

  Some of the K$^{-}$'s may escape from the nuclear medium, when they acquire enough energies from further collisions so as to be emitted as free K$^-$ mesons. This is a ``heating-up + escaping" process:
\begin{equation}
({\rm K}^{-}) _{medium}  \rightarrow ({\rm K}^{-}) _{heated}  \rightarrow ({\rm K}^-)_{free}.
\end{equation} 
 It is to be noted that the same attractive interaction is the origin of  \kbar~clusters. In this  sense, the ``subthreshold" K$^-$ mesons are brothers of \kbar~clusters; both are born from the same parents, {\it in-medium} K$^-$'s. \\

\noindent
{\it ii) Evolution of \kbar~clusters as deep trapping centers}

K$^{-} $'s may produce extra-deep and localized self-trapping potentials, as schematically shown in Fig.~\ref{fig:HI-continuum} (lower), which are intermittently accommodated by a few correlated nucleons (notably, p$^2$, p$^2$n ($^3$He) and p$^2$n$^2$ ($^4$He)). Under such circumstances, the K$^{-} $'s become self-trapped together with an ensemble of [ppn], for example. Since \kbar~clusters once produced are hardly destroyed by further collisions because of their extremely large binding energies compared  to the temperature, we expect a cascade evolution of \kbar~clusters, as shown below.\\

\begin{figure}[htb]
\vspace{-1cm}
\includegraphics[width=0.45\textwidth]{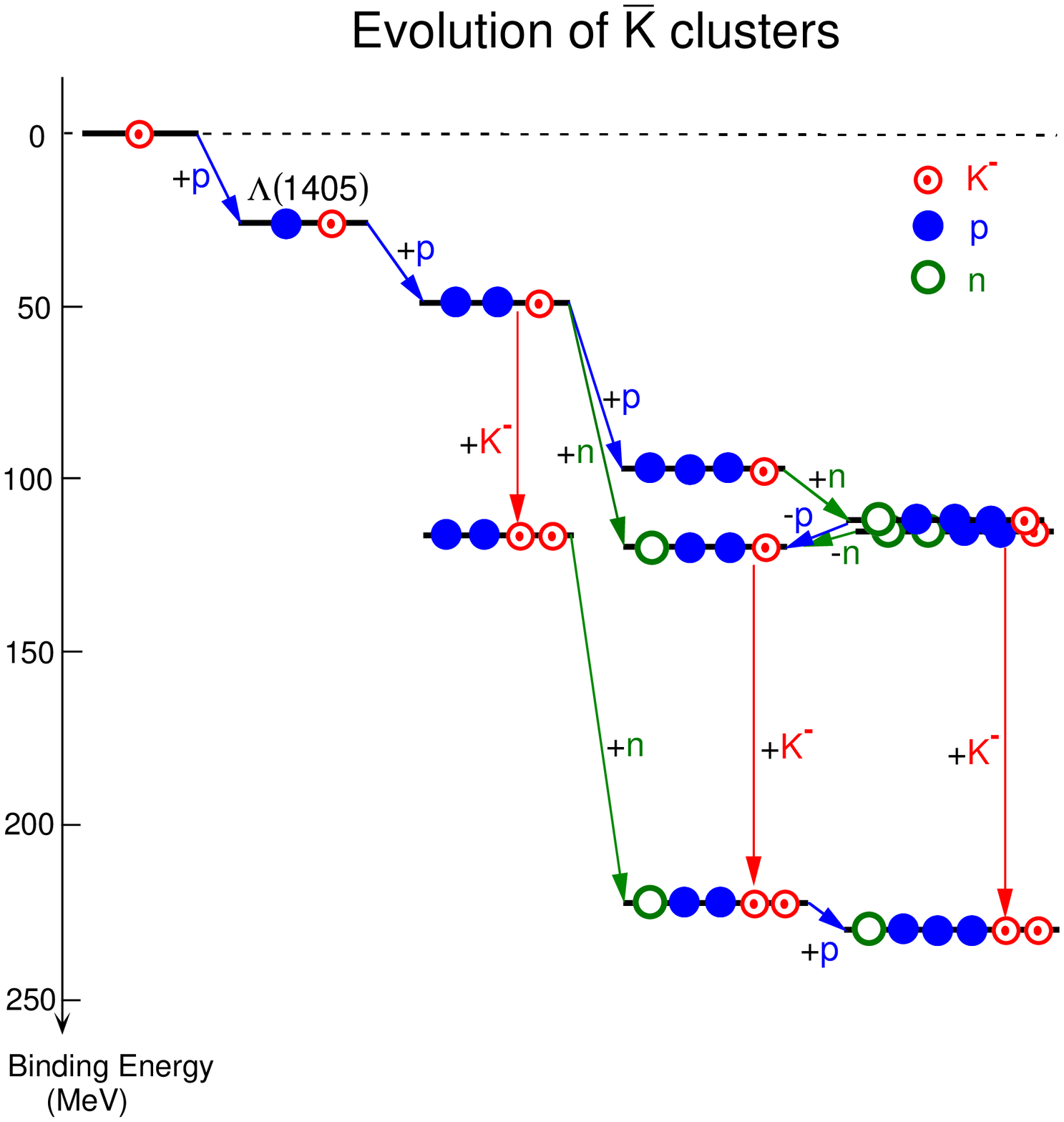}
\vspace{-1cm}
   \caption{\label{fig:evolution} Cascade evolution of \kbar~clusters as deep traps in heavy-ion collisions. The calculated bionding energies are shown.}
\end{figure}

\noindent
[Single-\kbar~cluster formation]
\begin{eqnarray}
p + K^- &\rightarrow& pK^-~(\Lambda_{1405})  \\
pK^- + p &\rightarrow& ppK^-  \\
ppK^- + p &\rightarrow& pppK^-  \\
ppK^- + n &\rightarrow& ppnK^-  \\
pppK^- + p &\rightarrow& ppppK^-  \\
pppK^- + n &\rightarrow& pppnK^-  \\
ppnK^- + p &\rightarrow& pppnK^-  \\
ppnK^- + n &\rightarrow& ppnnK^- \\
{\rm ^3 He} + K^- &\rightarrow&  ppnK^- \\
{\rm ^4 He} + K^- &\rightarrow&  ppnnK^- 
\end{eqnarray}
[Double-\kbar~cluster formation]
\begin{eqnarray}
ppK^- + K^- &\rightarrow& ppK^-K^-  \\
ppK^-K^- + n &\rightarrow& ppnK^-K^-  \\
ppnK^- + K^- &\rightarrow& ppnK^-K^-  \\
pppK^- + K^- &\rightarrow& pppK^-K^-  
\end{eqnarray}
These processes occur as {\it collisional capture processes}, when aided by surrounding nucleons, which transfer energies and momenta to form \kbar~clusters efficiently. 
The energy diagram for this cascade evolution was calculated, as shown in Fig.~\ref{fig:evolution}. The deepest trapping center among the single-\kbar~clusters is ppnK$^-$. The double-\kbar~clusters, ppnK$^-$K$^-$ and pppnK$^-$K$^-$, are the deepest among the double-\kbar~clusters. 
The probability of forming such deep traps can be estimated by a coalescence model \cite{Yazaki,Hirenzaki93}. Realistic simulations for heavy-ion reaction residues, such as RQMD \cite{Bass:98} and HSD \cite{Cassing:99}, can be extended so as to include the \kbar~cluster productions, which will be important.\\

\noindent
{\it iii) $\Lambda(1405)$ and $\Lambda(1520)$ as door-way particles}

Productions of $\Lambda(1405)$ and $\Lambda(1520)$ in heavy-ion reactions can also be sources of \kbar~clusters, since $\Lambda(1405)$ is a bound state of K$^-$ + p and $\Lambda(1520)$ is a resonance state of \kbar N. When they are produced in a nuclear medium, they proceed to kaonic bound states, forming \kbar~clusters. The role of these excited hyperons as doorways to kaonic systems was studied in the case of (K$^-,\pi^-$) reactions \cite{Yamazaki:02}. Likewise, excited hyperons with $S = -2$ can be a doorway to double-\kbar~clusters.\\

\noindent
{\it iv)  Direct formation of \kbar~clusters from QGP}

\begin{figure}[htb]
\includegraphics[width=0.5\textwidth]{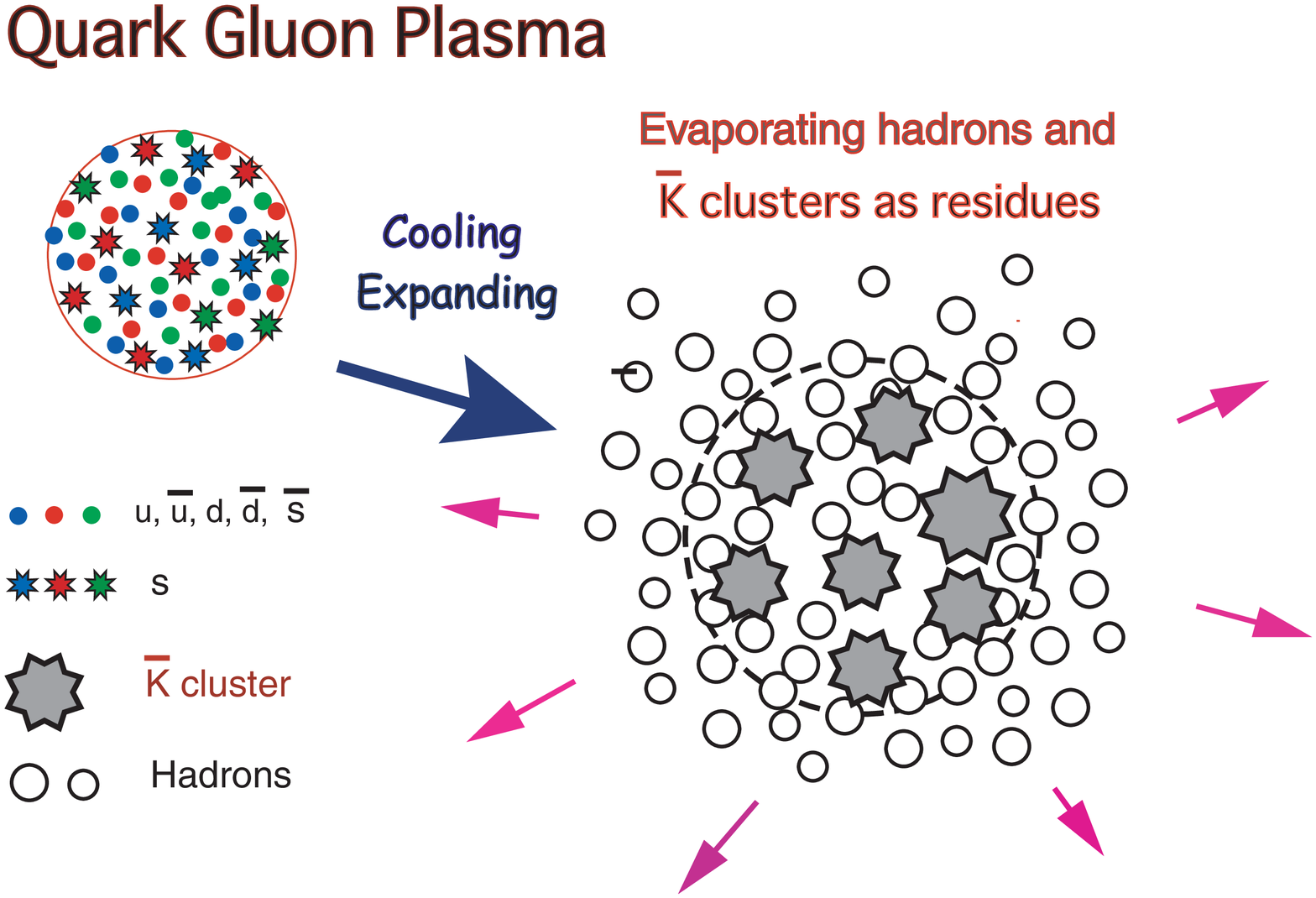}
   \caption{\label{fig:QGP} Quark gluon plasma and its transition to evaporating hadron gases with heavy and dense residues of \kbar~clusters. }
\end{figure}

When the temperature of a primordial fireball exceeds a freeze-out temperature ($T > T_f \sim 150$ MeV) it is expected to be in a hot quark-gluon plasma (QGP). Since the \kbar~clusters are by themselves dense,  and are likely to be in a deconfined quark-gluon phase, as in QGP, they will be spontaneously formed, like clusterized islands, remaining in an  expanding hadron gas medium throughout the freeze-out phase (see Fig.~\ref{fig:QGP}):
\begin{equation}
{\rm QGP} \rightarrow {\it evaporating~hadrons} + \overline{K}~{\it  clusters}.
\end{equation} 
Here, the $s$ quarks in a primordial QGP will act as seeds for  \kbar~clusters, which are eventually formed in a self-organized way and are decoupled from evaporating hadrons. In this way, \kbar~clusters are  produced directly as ``island-like" residues from QGP. This process is different from the cascade evolution process considered above, and the probability of each $s$-quark to proceed to a \kbar~cluster (even to a double-\kbar~cluster) is expected to be high. The time for their formation as well as the time for their decay are close to the freeze-out time.\\

\noindent
{\bf 4. \kbar~-cluster invariant-mass spectroscopy}\\

 Eventually, \kbar~clusters decay via strong interactions by their own intrinsic decay modes. Whether these decays occur inside or outside the nuclear collision volume, is a key problem. The condition to observe the free decay of a \kbar~cluster with a decay width $\Gamma_K$ is
\begin{equation}
\tau_K = \hbar/\Gamma_K > \tau_{f},
\end{equation} 
where $\tau_{f}$ is the freeze-out time. For $\Gamma_K = 20$ MeV, $\tau_K \sim 10$ fm/$c$, which is marginally longer than the calculated freeze-out time, $\tau_{f} \sim 5$ fm/$c$ \cite{Bass:98,Cassing:99,Friman:98,Li:00}. Thus, most \kbar~clusters formed in the freeze-out phase are likely to survive and undergo free decays. 

The above discussions indicate that the \kbar~clusters must be as abundantly produced as the free K$^-$ mesons, and that their decays can be tracked. 
The unique signature for \kbar~cluster formation is a clear peak to be revealed in the invariant-mass spectra of its decay particles,  
if all of the decay particles with their energies and momenta are correctly identified. This method applies to limited cases, where  \kbar~clusters can decay to trackable particles, such as
\begin{eqnarray}
&i)& ppK^- \rightarrow \Lambda + p,\\
&ii)& ppnK^- \rightarrow \Lambda + d,\\
&iii)& pppK^- \rightarrow \Lambda + p + p,\\
&iv)& ppnnK^- \rightarrow \Lambda + t,\\
&v)& pppnK^- \rightarrow \Lambda + {\rm {^3 He}},\\
&vi)& ppK^-K^- \rightarrow \Lambda + \Lambda,\\
&vii)& pppK^-K^- \rightarrow \Lambda + \Lambda + p,\\
&viii)& pppnK^-K^- \rightarrow \Lambda + \Lambda + d.
\label{eq:decay}
\end{eqnarray}
These decay processes are energetically the most favoured, though their branching ratios are not known. In the following, we show that this is indeed feasible. 

Recently, $\Lambda$ hyperons have been identified in high-energy heavy-ion reactions at GSI-SIS from the energies and momenta of their decay vertices, p + $\pi^-$, by a large 4$\pi$ detector (FOPI) \cite{FOPI}. The average multiplicity of $\Lambda$ at a H.I. energy of 2 $A$ GeV is about $0.15$ after a correction for the reconstruction efficiency, whereas the average multiplicity of p is about 40 \cite{FOPI}. Since the observed yields of d, t and $^3$He are also sizable, we expect that the formation of \kbar~clusters is highly probable. Invariant-mass spectra for the above processes can be composed from charged-particle tracks (p, d, t and $^3$He) in connection with a $\Lambda$-associated p - $\pi^-$ trajectories, though there will be a substantial background of combinatorial origin. 
The first goal should be to identify two important single-\kbar~clusters, ppK$^-$ and ppnK$^-$. The ppnK$^-$ nucleus is being searched for by using the $^4$He(stopped K$^-$, n) reaction at KEK \cite{Iwasaki:NIM}. 

Once single-\kbar~clusters are found, the next step will be to pursue double-\kbar~clusters. Abundant productions of $\Lambda$ are also observed at the RHIC energy  by PHENIX \cite{PHENIX} and STAR \cite{STAR}. Here, the multiplicities of K$^-$ are large, so that a large production of double-\kbar~clusters is expected. It is to be noted that the future GSI  accelerator will provide 40 GeV/u heavy-ions, which will be suitable for \kbar~cluster invariant-mass spectroscopy in view of the large baryon density to be achieved in collisions, and also of abundant strangeness production \cite{GSI-future}. \\

\begin{figure}[htb]
\includegraphics[width=0.4\textwidth]{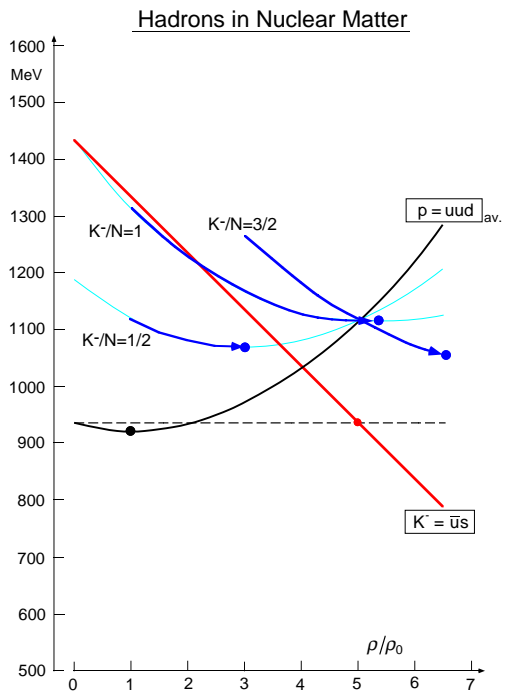}
   \caption{\label{fig:Kmatter} Schematic diagram for the density dependences of the bound-state energies of various baryon composite systems (pK$^-$)$^{m}$n$^{n}$. The \kbar N energy is represented by a red straight line, where no effect of chiral symmetry restoration is invoked. The nuclear compression is represented by a black curve and the total energies for representative fractions of K$^-$/N are depicted by blue curves.}
\end{figure}

\noindent
{\bf 5. Concluding remarks}\\

The present proposal is, in a sense, a revisit of an old proposal of {\it density isomers} by Lee and Wick \cite{Lee-Wick}. The time scale in the present case is around 10 fm/$c$, not ns - $\mu$s or longer, but the object is still {\it metastable} compared with the orbiting time of K$^-$ in nuclei ($\Gamma_K < E_K$). Another remark is that the invariant-mass method proposed here is a valid procedure because the decay of our concern takes place in free space after the freeze-out phase. On the other hand, invariant-mass spectroscopy for in-medium hadrons suffers from various  in-medium disturbances, as remarked in \cite{Yamazaki:99}. When \kbar~clusters decay in a dense fireball region, their ``invariant masses"  will be shifted to lower values ({\it collision-induced red shifts}) by an amount, $\Delta M \approx q^2/(2 M_N)$, as calculated in \cite{Yamazaki:99}. For $\rho \approx 3 \rho_0$ this ``red shift" amounts to  $\sim 50$ MeV.

We can conceive a further extension of the double-\kbar~systems to multi-\kbar~nuclear matter. Whereas the nucleons and hyperons are hard to compress, presumably because of the Pauli repulsion in the quark sector, multi-\kbar~systems, such as (pK$^-$)$^{m}$n$^{n}$, become self-compressed  dense matter without the aid of gravity. The characteristic feature of \kbar~in producing dense nuclear systems may be intuitively understood  as a result of the non-existence of Pauli blocking in the $(u, d)$ quark sector by implanting K$^-$, since K$^-$ is composed of $s\bar{u}$. Here, kaon condensation may also play an essential role \cite{Brown:94}.  Fig.~\ref{fig:Kmatter} shows schematically the expected dependences of multi-\kbar~bound states as compared with non-\kbar~nuclei. The \kbar~matter with a large \kbar~fraction (K$^-$/N $\sim$ 1) may be more stable than the corresponding non-strange matter.
The equation of state for describing {\it gravity-assisted} dense stars will be obtained from empirical bases, when the double-\kbar~nuclei, as predicted here, are investigated experimentally. 

So far, the present treatment does not contain the effect of chiral symmetry restoration at high density. If the $\bar{\rm K}$N interaction is increased along with a restoration of the chiral symmetry, as observed in deeply bound pionic nuclei \cite{suzuki:02}, the K$^-$ energy line is bent downward with the increase of $\rho$; \kbar~clusters may be more bound and denser, and the \kbar~matter may become more stable. The \kbar~(or $s$-quark) clusters, which we propose to study experimentally, will provide not only a unique playground to study possible quark-gluon phases of dense and bound nuclear systems, but also a key to understanding the neutron and strange matter. \\

\noindent   
{\bf Acknowledgement}\\

We would like to thank Professor Paul Kienle for his stimulating and encouraging comments. We also thank Professors H. Horiuchi, N. Herrmann and R.S. Hayano, and Dr. K. Iida for the helpful discussion. The present work is supported by Grant-in-Aid of Monbukagakusho of Japan. One of the authors (A. D.) acknowledges the receipt of a JSPS Postdoctoral Fellowship.

\end{document}